\documentclass[a4paper,11pt]{article}
\usepackage{jheppub} 
\usepackage[T1]{fontenc} 
\usepackage{hyperref}
\usepackage{tikz}
\def\beq{\begin{equation}}
\def\eeq{\end{equation}}
\def\C{\mathcal C}

\title{\boldmath  Future Boundaries and the Black Hole Information Paradox.}

\author[a,b,c]{Malcolm J. Perry }
\affiliation[a]{School of Physics and Astronomy, Queen Mary University of London, Mile End Road, London E1 4NS, UK.}
\affiliation[b]{DAMTP, Centre for Mathematical Sciences, Wilberforce Road, Cambridge, CB3 0WA, UK.}
\affiliation[c]{Trinity College, Cambridge, CB2 1TQ, UK.}
\emailAdd{malcolm@maths.cam.ac.uk}

\abstract{The black hole information paradox is the incompatibility of quantum mechanics with the 
semi-classical picture of Hawking radiation. Hawking radiation appears thermal and eventually leads  
to the complete disappearance of a black hole. However, black holes could be formed from a pure quantum
state. The transition from such an initial state to the final state of pure Hawking radiation cannot
be described by  unitary time evolution. In this paper, we present an analysis in quantum gravity that 
shows how boundary conditions in the future prevent a loss of quantum mechanical information from the
spacetime. In classical physics, the future boundary of the spacetime in the black hole interior
is a singularity. Realistic gravitational collapse results in a BKL type of approach to the singularity.
But, solving the Wheeler-DeWitt equation reveals that the singularity does not form and can be replaced
by specifying a final state density matrix. Such a condition is natural within the context of  
consistent histories version of quantum mechanics. We provide a self-contained treatment of these issues.
How information escapes from the black hole will be treated elsewhere. 
\footnote{If I have omitted a reference to relevant work, please let me know 
and I will incorporate it into this paper in its final form.} }

\begin{document}

\maketitle
\flushbottom

\section{Introduction}
\label{sec:intro}

The information paradox has been with us for almost fifty years now, following on from Hawking's discovery
\cite{Hawking:1974rv}, \cite{Hawking:1974sw}  that black holes appear to radiate thermally at a temperature
$T_H$.
The first law of black hole mechanics \cite{Bardeen:1973gs} expresses the change in 
the mass of a black hole in an 
asymptotically flat spacetime as its 
angular momentum and electric charge change infinitesimally when a black hole transitions between two 
equilibrium states.
In such a process, 
\beq
dM = \frac{\kappa}{8\pi}dA + {\bf \Omega}\cdot d{\bf J} + \Phi\, dQ \label{eq:1stlaw}
\eeq
where $M$ is the mass of the black hole, $\kappa$ its surface gravity, $A$ the area of the event horizon, 
${\bf \Omega}$ its angular velocity, ${\bf  J}$ its angular momentum, $\Phi$ its electrostatic potential
and $Q$ its electric charge \footnote{We  use natural units such that
$G=c=\hbar=k=1.$ The signature of spacetime is taken to be $(-+++)$. The Riemann tensor is defined 
by $(\nabla_a\nabla_b - \nabla_b\nabla_a) V_c = R_{abc}{}^d V_d.$}. The first law 
is a theorem in classical general relativity for black holes in an asymptotically flat spacetime.
The first law also reflects the no-hair theorems which tell us that the geometry of a stationary 
black hole, is completely described  by $M, \,{\bf J}$ and $Q$ and that the metric is given by the 
Kerr-Newman solutions, 
\cite{Schwarzschild:1916uq,Reissner:1916zz,Weyl:1917gp,Nordstrom:1918zz,Kerr:1963ud,Newman:1965my}.
For a recent review of the uniqueness theorems, 
see \cite{Chrusciel:2012jk}. 

The temperature of Hawking radiation, $T_H$  is given by
\beq
T_H = \frac{\kappa}{2\pi}
\eeq
which allowed a reinterpretation of the first law of black hole mechanics as the first 
law of thermodynamics \cite{Hawking:1974sw}
and led to the identification of black hole entropy $s_H$ in terms of the horizon area as
\beq
s_{H}=\frac{A}{4}.
\eeq

The idea that area of the event horizon could be identified with 
entropy had previously been suggested by Bekenstein, 
\cite{Bekenstein:1972tm,Bekenstein:1973ur,Bekenstein:1974ax}.
His arguments were based on an examination of the laws of thermodynamics for a black hole
interacting with matter. He argued that if the second law of thermodynamics for such a system was
true, then the black hole must have an entropy proportional to the area of the horizon. 
However, he was unable to fix the constant of proportionality. We will refer to $s_H$ as 
the Bekenstein-Hawking entropy.

The identification of the horizon area as black hole entropy is supported by another result in 
classical general relativity, the area theorem.
Any change to a black hole that results in a change of its horizon area by $\Delta A$ is such
that
\beq
\Delta A \ge 0 
\eeq
provided any matter involved obeys the weak energy 
condition, \cite{Hawking:1971tu,Hawking:1971vc}. The identification of the horizon area
as entropy means that the area theorem can be reinterpreted as the second law of thermodynamics
\footnote{Recently, the data from the black hole merger GW150914 was examined with the aim
 of testing the area theorem. Agreement with the area theorem was found with $97\%$ probability 
\cite{Isi:2020tac}.}. 

Since black holes  produce thermal radiation, they will
eventually disappear. Suppose a black hole has zero angular momentum and zero electric charge, the
Hawking temperature is then
$T_H=1/(8\pi M)$. As it produces Hawking radiation, it will lose mass and therefore get hotter.
It will radiate away all its mass on a timescale $\tau \sim M^3$.
Eventually, the black hole will disappear leaving nothing but Hawking radiation. 

Another view of black hole entropy is provided by a calculation of Gibbons and 
Hawking \cite{Gibbons:1976ue}. They calculated the partition function of stationary black holes
using Euclidean field theory techniques. Euclidean spaces that are the analytic continuation 
of stationary black holes exhibit a periodicity in imaginary time corresponding to the inverse
temperature. Under Euclideanization, the horizon becomes a conical singularity unless this periodicity
is imposed. There is no analogue of the interior of the black hole in this picture
so in effect one has traced over the interior states of the black 
hole, {\cite{Hartle:1976tp,Gibbons:1976pt}.  Starting from the path integral for gravity 
using the Einstein action, Gibbons and Hawking found that the thermodynamic entropy
of the black hole was again given by $A/4$.

Now  consider black hole formation in classical physics.
Some matter undergoes gravitational collapse and once it has become sufficiently compressed, 
an event horizon is expected 
to form and then a spacetime singularity  will develop. An explicit example of how this comes
about was provided by Oppenheimer and Snyder \cite{Oppenheimer:1939ue} for the case of dust 
undergoing spherically symmetric collapse.
The interior of the dust cloud, mass $M$, is described by a contracting 
Friedmann-Robertson-Walker universe, whilst 
exterior to the body, the spacetime is that given by the Schwarzschild metric. Once the outer surface of 
the collapsing cloud passes through the Schwarzschild radius, $r=2M$, an event horizon forms. The end 
point of collapse is a spacelike singularity.

Gravitational collapse is not usually 
spherically symmetric.
Nevertheless, this general picture is expected to still hold. The hoop conjecture \cite{Thorne:1972ji} 
suggests
that once matter is sufficiently condensed, an horizon will 
form, \cite{Schoen:1983zz, Gibbons:2009xm, Cvetic:2011vt}. There is at present no complete proof 
of the hoop conjecture, but it is widely believed and is supported by numerical
evidence \cite{Cardoso:2014uka}. Penrose's singularity 
theorem \cite{Penrose:1964wq}
guarantees that a singularity will form once an horizon has appeared. 
The weak cosmic censorship conjecture supposes that the singularity will be 
hidden from observers in the asymptotic region. Stated rather more precisely, it requires that 
in an asymptotically flat spacetime,
future null infinity should be 
geodesically complete, \cite{Penrose:1969pc,Geroch:1979uc}.
Again, there is no complete proof of the weak cosmic censorship conjecture but again 
it is widely believed to be true and there is numerical evidence 
in its favour  \cite{Cardoso:2014uka}.
If a black hole has rotation or electric charge 
and is stationary,  its geometry is that of the Kerr-Newman solution and has an inner Cauchy horizon.
The inner horizon has been shown to be unstable perturbatively \cite{Simpson:1973ua} and it is 
believed that in 
the exact theory, it will morph into a singularity. 
Unambiguous evidence for this has been provided by Dafermos and Luk 
\cite{Dafermos:2017dbw}  who showed that in certain circumstances, a singularity forms in 
place of the Cauchy horizon. 
In general, singularities that replace a null Cauchy horizon are expected to be spacelike 
although it is possible that there are
null segments. What seems to be ruled out in realistic situations, is the possibility 
of the timelike
singularities of the type that are found in the maximal analytic extension of the stationary solutions
in the Kerr-Newman family. As a consequence of all of this, we expect
the Penrose diagram of an evaporating black hole
to be as shown in Figure $1$.

Now let us ask  what happens quantum mechanically. Our only option is to try to perform a  
semi-classical treatment since, as yet, there is no satisfactory theory of quantum gravity. 
In Figure $1$, $\Sigma_i$ is a surface on which an initial state can be specified,
$\Sigma_f$ is a final surface outside the black hole after it has completely evaporated 
and where a final state can be described.
Lastly, $\Sigma_s$ is a surface close to the singularity and anything crossing this surface 
would seem to impact the singularity and be lost to spacetime.
Quantum mechanics tells us that, in the Heisenberg picture, one 
expects there to be a unitary operator  $U(t)=e^{iHt}$ with $H$ being the Hamiltonian
such that any operator ${\cal O}$ evolves with time by
\beq
{\cal O}(t^\prime) = U(t^\prime-t)\ {\cal O}(t)\ U(t-t^\prime),
\eeq
whereas states remain constant.
The density matrix $\rho$ therefore remains constant and hence the von Neumann, or entanglement, 
entropy \cite{Neumann:1932} $S$, given by
\beq
S = - tr\, \rho\, \ln \rho
\eeq
also remains constant. In particular, the entropy on the surface $\Sigma_i$ should be the same as 
the entropy on $\Sigma_f$.  
The initial state could have been pure with 
zero von Neumann entropy. However, the semi-classical picture appears to tell us that after 
the black hole has disappeared, there is 
only thermal Hawking radiation and that has non-vanishing entropy. Such evolution cannot be described by 
quantum mechanics as the entropy is not constant. 
This is the information paradox. It has come about because the matter 
falling into the black hole encounters
a boundary to spacetime at the singularity. At this boundary, at least classically, one presumes the 
matter has left the spacetime.

Initially, Hawking suggested that quantum mechanics breaks down \cite{Hawking:1976ra}.
That is certainly a logical possibility.
However, these ideas were criticised by Gross \cite{Gross:1983mq},
Banks, Susskind and Banks \cite{Banks:1983by} and Lee \cite{Lee:1985rp}. 
Despite such criticism, 
a recent review by Unruh and Wald \cite{Unruh:2017uaw} suggests that we somehow have to 
learn to live with information loss. There have been various other suggestions. There
might be black hole remnants that have large entropy and whose states provide for the purification
of the Hawking radiation.
However, any such remnant would have to be small as there is no reason for the Hawking radiation to turn 
off until the black hole is of roughly the Planck scale where non-perturbative quantum gravity effects 
might take place and allow this to happen. However, for something small to contain a large 
amount of information appears unnatural. It might be that instead of a singularity, 
there is a baby universe that splits off
and forms a disjoint part of spacetime from the original one. In either of these cases
information will be lost as far as observers in the asymptotic region are concerned. For a review of
this type of possibility, see \cite{Polchinski:2016hrw}.
Another possibility is that black holes never form, but instead one forms a fuzzball, a region
 of spacetime
that at large distances looks like a black hole but does not contain either an horizon or a 
singularity \cite{Mathur:2005zp}. 
These would be quantum gravity configurations that have no classical analogue as classical 
general relativity does not admit such solutions \footnote{At least not in dimension four.}. 
Furthermore, for very large black holes as found 
at the centers of galaxies, one expects fields outside the horizon to be very weak. One would
not expect quantum gravity effects to be important at the horizon and so it is hard to 
see how in the fuzzball picture an horizon would fail to form.  

The AdS-CFT correspondence \cite{Maldacena:1997re} seems to indicate that quantum mechanics 
should work as expected 
for a description of black hole evaporation.  That is because one can have evaporating black holes 
in a spacetime that is asymptotic to anti-de Sitter space. The boundary theory that is dual to
the bulk theory in anti-de Sitter space is perfectly unitary and defined in a way that is non-perturbative,
\cite{Maldacena:2001kr}.
The puzzle then is to determine where the semi-classical picture is wrong. Whatever the problem is, 
one should be able to rely on semi-classical arguments as long as fields do not approach Planckian
scales. That seems to indicate that quantum gravity effects are important but only as one 
gets close to the singularity. 

This paper is an attempt to understand what properties of quantum gravity
allow a resolution of the information paradox. We require there to be the standard type of quantum 
mechanical evolution for observers exterior to the black hole. 
What then are the implications of requiring the evolution 
from $\Sigma_i$ to $\Sigma_f$ to be unitary?
For some relatively recent reviews of the information paradox see Mathur
\cite{Mathur:2009hf} and Harlow \cite{Harlow:2014yka}. See also lecture six in
\cite{Aaronson:2016vto} for an information theoretic view of the problem.

The primary obstacle is the lack of a microscopic, or fine-grained, theory of quantum gravity.
Such a theory would allow a calculation of the complete details of how a black hole evaporates.  
We do have a coarse-grained description based on the geometrical picture provided by general relativity
but this a classical theory and does not contain a description of the quantum phenomenon of 
Hawking radiation. We do have a semi-classical picture based on the path integral quantisation 
of general relativity. The failure of renormalizability limits its usefulness \cite{tHooft:1974toh}}.
This too is a geometrical picture but its rules seem to imply that a strict interpretation based on
smooth, real, Lorentzian metrics needs to be abandoned. At the very least, it requires us to veer off
into the complex and it does not really allow us to develop a coherent approach to singularities.
The path integral is however equivalent to the canonical approach \cite{Leutwyler:1964wn,Faddeev:1973zb,
Fradkin:1977hw,Hartle:1983ai,Barvinsky:1986wi}
\footnote{These references show 
the equivalence in a 
rather formal way and the treatment ignores issues of the spaces on which the path integral and
the Wheeler-DeWitt equation are defined, 
operator ordering and ultraviolet divergences and therefore
can be relied on only in the WKB approximation.}.
In the canonical approach, 
the Wheeler-DeWitt equation \cite{Wheeler:1967xx, DeWitt:1967yk}
defines the wavefunction of the universe, $\Psi[\gamma]$ as a functional on superspace. 
Superspace is the space of all $3$-metrics modulo diffeomorphisms. 
Although $\Psi[\gamma]$ has as its argument the $3$-metric $\gamma_{ij}$, 
it depends only on the geometry of the spacelike surface on which it is defined
and not on the coordinate system chosen to define $\gamma_{ij}$.
Superspace includes the possibility that the $3$-metric is singular and therefore 
there is the possibility that the wavefunction of the universe has something to say about singularities.
In classical physics, it appears that the approach to singularities is chaotic as was first
discussed by Belinsky, Khalatnikov and I.~M.~Lifshitz (BKL), 
\cite{Belinsky:1970ew,Belinsky:1982pk,Belinsky:aa}. Their method was based on the canonical approach 
to general relativity. As such, it lends itself to an extension to a quantum version by an appropriate
adaptation of the Wheeler-DeWitt equation. A recent discussion of such methods as applied to the origin of 
the universe can be found in \cite{Nicolai:2021raw}.
Although $\Psi[\gamma]$ was originally defined by the Wheeler-DeWitt equation, it can also be found
by path integral methods \cite{Hartle:1983ai}.

There are two principal difficulties posed by the black hole information paradox that require
resolution by a fine-grained theory of quantum gravity.  The first is that 
the interior of the black hole is causally disconnected from any external observer.  
There seems to be  no way to recover material that falls into the black hole. 
The second is that the singularity 
simply swallows everything up, taking it beyond the boundary of spacetime.  

In his work on cosmology, Hawking asserts that \lq\lq There ought to be something very
special about the boundary conditions of the universe and what can be more special than 
the condition that there is no boundary,\rq\rq\ \cite{Hawking:1981gb}. 
The laws of physics are CPT-invariant. So perhaps 
this provides us with a clue about how to proceed. We simply need to find what boundary conditions apply 
to the singularity inside the black hole. That one might need to consider boundary
conditions in the future in quantum mechanics was first suggested by 
Einstein, Tolman and Podolsky \cite{Einstein:1931zz}.  Aharonov, Bergmann and Lebowitz \cite{Aharonov:xz}
presented a formulation of quantum theory that incorporated the possibility of fixing
boundary conditions in the future. Subsequently, their scheme was elaborated on by Gell-Mann and Hartle,
\cite{GellMann:1991ck}.

Horowitz and Maldacena \cite{Horowitz:2003he}
made the suggestion that such methods may be used to explore and perhaps resolve the black 
hole information paradox.
However,
difficulties with these proposals were pointed out by Gottesmann and Preskill \cite{Gottesman:2003up} 
and by Bousso and Stanford \cite{Bousso:2013uka}. Subsequently,
Lloyd  \cite{Lloyd:2004wn} and by Lloyd and Preskill \cite{Lloyd:2013bza} discussed these difficulties
and showed how they could be resolved. But, in the absence of any 
concrete proposals about how to use such techniques, interest appears to have waned. 

 In this paper, we present a calculation that shows that there is a boundary condition for the 
 singularity that emerges in a natural way from the coarse-grained semi-classical picture.
 By examining solutions of the Wheeler-DeWitt equation, we are able to show that singularities
 in gravitational collapse simply do not form.  The probability amplitude
 for geometries on $\Sigma_s$, as shown in Figure $1$, vanishes in the limit that $\Sigma_s$
 tends towards the classical singularity.  We conclude  
 that there is no singularity. Nevertheless,  in the above
 limit, $\Sigma_s$  presumably
 still represents a future boundary to the spacetime. One can impose a boundary condition on $\Sigma_s$
 so that 
 information does not escape from the spacetime. 
 We have provided half of a solution to the information paradox. It is
 still necessary to address the issue of how the information emerges from 
 the black hole. That problem will be addressed elsewhere.

 The plan of the remainder of the paper is to explain how one necessarily arrives at our conclusion.
 We describe the results that motivate our treatment and briefly outline the tools required
 to see how everything fits together. Our aim being to make the work self-contained.
 In section two, we describe the Page curve and how it arises.  In section three we 
 discuss the issues raised by firewalls and the possibility of their avoidance due to 
 appearance of extra gravitational contributions to the fine-grained black hole entropy. 
 In section four, we describe the consistent histories approach
 to quantum mechanics and how it can implement a coarse-graining on a fundamental quantum theory. 
 In section five, we discuss the time symmetric version of quantum theory and post-selection. 
 In section six, we examine the formation of singularities in gravitational collapse and the BKL
 phenomenon. In section seven we apply the ideas of quantum gravity to see what the boundary 
 condition must be. In the last section, we first briefly examine the effect of matter
 contributions to our picture, and the possible extension to dimensions higher than four.  
 We conclude by looking at some of the unconventional phenomena that one
 might expect to find inside the horizon and then move on to discuss 
 possible observable consequences. For strange behaviour not to be observable outside the horizon,
 we propose a quantum cosmic censorship conjecture.  A brief outline of the work presented in this paper 
 is to be found in
 \cite{Perry:2021mch}.

\section{Black Hole Evaporation and the Page Curve}
\label{sec:BHPC}

In Hawking's original picture, a Schwarzschild black hole will produce black body radiation and thereby
lose mass at a 
rate
\beq \dot M = -\sigma AT_H^4 \label{eq:loss} \eeq
where $\sigma$ is a number of order unity that depends on the particle species being radiated, 
the Stefan-Boltzmann constant and the grey-body factors of the black hole. Its
precise value does not concern us here. 
Integrating (\ref{eq:loss}) we find the mass of the black hole as a function of time, $M(t)$.  
Taking the mass to be $M_0$ 
at time $t=0$  and neglecting any time dependence in $\sigma$, $M(t)$ is given by 
\beq M(t) = \Biggl(M_0^3 - \frac{3\sigma t}{256\pi^3}\Biggr)^{1/3} \label{eq:mass} \eeq
The lifetime of the black hole $t_l$ is found from (\ref{eq:mass}) giving
\beq t_l = \frac{256\pi^3M_0^3}{3\sigma}. \eeq 
Assuming the Hawking radiation to be thermal,
the entropy of the radiation $s_r(t)$,\footnote{We reserve lower case $s$ for the thermodynamic, 
or coarse-grained entropy and upper case $S$ for the von Neumann or entanglement entropy.}
is generated at a rate
\beq \dot s_r = \frac{4}{3}\sigma AT_H^3. \eeq
After a time $t$, the entropy is 
\beq s_r(t) = \frac{16\pi}{3}\Biggr[M_0^2 - \Biggl(M_0^3 - \frac{3\sigma t}{256\pi^3}\Biggr)^{2/3}\Biggr].\eeq
We note that when the black hole has completely disappeared at time $t_l$, the entropy 
in the radiation is 
\beq
s_r(t_l) = \frac{16\pi M_0^2}{3}. \label{eq:radent} \eeq
$s_r(t_l)$ is greater than the Bekenstein-Hawking entropy $4\pi M_0^2$ of the black hole 
when it is first formed.
The generalised second law of thermodynamics 
supposes that the total thermodynamic entropy of matter together with that of the black hole to be an
increasing function of time, \cite{Bekenstein:1972tm,Bekenstein:1974ax} 
and is supported by (\ref{eq:radent}).

Page has presented some more refined versions of this 
calculation in \cite{Page:1976df,Page:1976ki,Page:1983ug}. He included the black hole grey-body factors,
and the temperature dependence in $\sigma$ coming from different particle species. His conclusions 
are in accord with our crude estimate presented above.

This result is at the heart of the information paradox. If quantum mechanics applies to an evaporating 
black hole and the black hole was formed from the collapse of some matter in a pure quantum state, 
the final von Neumann entropy should be zero. The conclusion is therefore that our treatment so far has 
included some kind of hidden coarse-graining.  A true microscopic treatment would reveal where and 
how this has taken place. The difficulty is to see how this could have happened. The black hole
uniqueness theorems appear to indicate that information about the collapse that formed 
the black hole cannot reside in the geometry outside the black hole. Soft hair provides 
for the possibility that some extra information beyond the mass, angular momentum and electric charge
could be detectable outside the black hole \cite{Hawking:2016msc}, but this does not appear 
to be sufficient to resolve
the issue. Then again, communication from the interior of the black hole to the exterior 
would violate our ideas of causality. So on the face of it, one appears stuck.

It is generally presumed that conventional quantum mechanics works to describe physics
outside a black hole. We now make some proposals as to how this can come about. Before we proceed, 
we need to be clear about what assumptions we are making. 
Firstly, we take it that the black hole behaves as a conventional quantum mechanical system as 
seen by asymptotic observers. The black hole density of states is given by $e^{s_H}$ with $s_H$ being the 
Bekenstein-Hawking entropy. Secondly, interaction of the black hole with its environment is determined by 
unitary time evolution.
These assumptions have been termed the \lq\lq central dogma\rq\rq\ by
Almheiri, Hartmann, Maldacena, Shaghoulian and Tajdini \cite{Almheiri:2020cfm}.

We now take these assumptions as controlling the physics of black hole evaporation. 
Working in the Heisenberg picture, we can start with the collapsing matter being described by
a density matrix $\rho$ with von Neumann, or entanglement, entropy $S$. 
Once a  black hole has formed, we can divide the system into two parts, $r$ being the 
radiation with density matrix $\rho_r$ and the black hole $h$ with density matrix $\rho_h$. 
Eventually, we will be left just with radiation which is again described by the density
matrix $\rho$ assuming that evolution really is unitary.
Page examined such a system \cite{Page:1993df,Page:1993wv,Page:2013dx} 
and we very briefly review his result here.  
The reduced density matrices of the radiation $\rho_r$ and the black hole $\rho_h$
are given by
\beq \rho_r = tr_h\, \rho \ \ \ \ \ \ \rho_h = tr_r\, \rho \eeq
where $tr_r$ and $tr_h$ are traces taken over the degrees of freedom of the radiation and black hole 
sectors respectively. These density matrices can be used to compute the von Neumann entropy of 
the radiation $S_r$  or the black hole $S_h$,
\beq S_r = -tr_r\, (\rho_r \ln \rho_r) \ \ \ \ \ \ S_h = -tr_h\, (\rho_h \ln \rho_h). \eeq
Let $S$ be the entropy of the combined system $r\, \cup \, h$. Then $S,S_r$ and $S_h$ obey 
the subadditivity and Araki-Lieb inequalities for
a composite system with two components \cite{Nielsen:xx}, 
\beq S_r + S_h \ge S \ge \vert S_r - S_h \vert. \eeq 
This pair of inequalities are often referred to as the triangle inequalities.  

Suppose that initially the collapsing matter is in a pure state so that $S=0$. Then $S_r=S_h$.
Contrast $S_r$ and $S_h$ with the thermodynamic, or coarse-grained, entropy $s_r$ and $s_h$. 
They display marked differences to each other as $s_r \ne s_h$. The 
Bekenstein-Hawking coarse-grained entropy of the black hole is initially extremely large. 
Similarly, the coarse-grained entropy of the radiation is low simply because the temperature is low. 
At late times, the Bekenstein-Hawking entropy is low because the black hole has become small.
But the total entropy in the radiation is large as can be seen from (\ref{eq:radent}).

Page's insight follows from the \lq\lq central dogma.\rq\rq \ The black hole is treated as conventional 
quantum system described by a Hilbert space of dimension $n$. The radiation is a conventional 
quantum system described by
a Hilbert space of dimension $m$. The dimensionality of the combined Hilbert space is thus $nm$. 
Suppose the system is initially in a pure state $\vert x \rangle$ so that its density matrix is
$\rho_x=\vert x \rangle \langle x \vert.$ At the beginning of the evaporation process, it is assumed 
that the state of the radiation is described by picking states at random out of the $nm$ possible
states of the combined system. This amounts to selecting $m$ random Hermitian matrices $M_\alpha,
\ \ \alpha=1, \ldots m$.  In a diagonal basis this results in a density matrix for the radiation
\begin{equation} \rho_r = \sum_{\alpha=1}^m\ p_{\alpha i}\
\vert i \rangle \langle i \vert, \ \ i=1, \ldots mn
\end{equation}
where $p_{\alpha i}$ are the eigenvalues of $M_{\alpha}$. The probability $P$ of finding the density
matrix $\rho_r$ given the initial density matrix is $\rho_x$ is given by
\begin{equation} P  = {\rm Tr}\ (\rho_r\rho_x) = \sum_{\alpha i} p_{\alpha i}\ \langle i \vert x \rangle \,
\langle x \vert i \rangle = \sum_\alpha\ p_{\alpha x}. \end{equation}
The von Neumann entropy of $\rho_r$ is 
\begin{equation}
S_r = - \sum_\alpha\  \langle p_{\alpha x} \ln p_{\alpha x} \rangle. \end{equation} 
The expectation value can be computed using random matrix techniques 
\cite{Page:1993df,Page:1993wv,Page:2013dx,
Wigner:1959,Mehta:xx,Dyson:1962brm,Lubkin:xx,Lubkin2:xx,
Lloyd:1988cn}.
For the case $m\ll n$ corresponding to early times, one finds 
\begin{equation} S_r \sim \ln m -\frac{m}{2n} \label{eq:ent} \end{equation}
The entropy in the radiation is close to its maximal value $\ln m$, since $m\ll n$. 
The quantum information in the radiation therefore starts off very low. $S_r$ is an increasing function of
$m$ until $m\sim n$. 

At late times, most of the system is Hawking radiation and so $m\gg n$. Now we can regard the state of 
the black hole as being modelled by choosing $n$ random matrices. Then we can find $S_h$ by the same method
but since $m\gg n$ all we need to do to find $S_h$ is to interchange $n$ and 
$m$ in (\ref{eq:ent}). Since $S_r=S_h$ and we get
\begin{equation} S_r  \sim \ln n - \frac{n}{2m}. \end{equation}  
Now $n$ is a {\it decreasing} function of time, so $S_r$ is decreasing and therefore contains a 
lot of quantum information. 

The quantum  information content of the radiation 
starts off low but starts to increase dramatically from a time where $m \sim n$, the Page time,
$t_p$. Page showed that
\beq t_p \sim  \Bigl(1-\frac{8}{5^{3/2}}\Bigr)t_0 \sim 0.28 t_0 \eeq
The Page time $t_0\sim M_0^3$. The black hole does not start to allow information to appear
in the radiation at some specific mass but rather at 
around a quarter of the lifetime  of the black hole. The indications therefore are 
not that Planck scale quantum gravity effects
change the evaporation process once the black hole has reached some mass threshold but rather
information starts to leak out in a continuous fashion right from the outset being at first very gradual 
but later much more rapid. A black hole is said to be young before the Page time and old after it.  For
young black holes the entropy in the radiation is increasing but for old black holes the 
entropy in the  radiation is decreasing. In addition, it means that the radiation from old black 
holes must be maximally entangled with that from young black holes.

 \section{Firewalls}
 \label{sec:Firewalls}
 
 The conventional picture of  Hawking radiation is that it is the result of the production of 
 particle
 anti-particle pairs
 close to the horizon.
 A particle escapes to infinity and its anti-particle partner falls into 
 the black hole. However, there is a serious difficulty with this picture if unitary evolution 
 of the black hole is to be retained, 
 \cite{Mathur:2009hf,Braunstein:2009myy}.
 
 Consider an old black hole; the existing Hawking radiation is in a state $r$ with density 
 matrix $\rho_r$
 and the black hole is in a state $h$ with density matrix $\rho_h$. The black hole then evolves for 
 a short interval of time,
 creating more Hawking radiation in a state $y$. This radiation
 escapes to infinity and has density matrix $\rho_y$. The new composite system of radiation 
 is $r^\prime$ with density matrix $\rho_{r^\prime}$. The anti-particles falling into the black hole
 are in a state $\bar y$, density matrix $\rho_{\bar y}$ and, being pair-produced with $y$, are 
 maximally entangled with the
 particles that were radiated off to infinity. After this has happened, the black hole is in a new state
 $h^\prime$ with density matrix $\rho_{h^\prime}$. To each of these states, one can assign a von Neumann 
 entropy. Since $y$ and $\bar y$ are maximally entangled, $S_y=S_{\bar y}$. $r^\prime$ is a composite state
 formed from $r$ and $y$, so that $r^\prime=r\otimes y$ and $\rho_{r^\prime}=\rho_{ry}$. In the same way,
 the composite state $y\otimes \bar y$ has density matrix $\rho_{y\bar y}$. However, since $y$ and 
 $\bar y$ are maximally entangled, $S_{y\bar y}=0$. Similarly $h^\prime=h \otimes \bar y$, has density
 matrix $\rho_{h\bar y}$ and von Neumann entropy $S_{h\bar y}$.
  
 Strong subadditivity of entropy is an inequality that applies to
 a combination of three quantum systems, \cite{Nielsen:xx}. Provided conventional quantum mechanical
 properties hold for the states in question, then if one takes any three quantum systems $a,b$ and $c$,
 the entanglement entropies of the composite states $ab$ and $ac$ obey
 $S_{ab}+S_{bc} \ge S_a + S_c $. Now putting $a=r$, $b=y$ and $c=\bar y$
 results in
 $S_{ry} + S_{y\bar y} \ge S_r +S_{\bar y}$ which can be re-arranged to give 
 $S_{ry}- S_r - S_{\bar y} \ge 0$ since $S_{y\bar y}=0$. But for old black holes, the von Neumann 
 entropy of the
 Hawking radiation is decreasing so that
 $S_{ry}\le S_r$ and we reach an inconsistency.
 Essentially, the strong subadditivity inequality says that if there is some unitary process in which
 a particle escapes observation, in this case by disappearing into the black hole, then the
 entropy of the remainder of the system must increase. However, the total von Neumann 
 entropy in radiation from old 
 black holes is decreasing as is required by unitary time evolution. We conclude that the overall 
 picture is inconsistent. The use of strong subadditivity, or perhaps the assumptions under which strong subadditivty was 
derived, are suspect \footnote{One is reminded of the aphorism due 
to Arthur Eddington. \lq\lq If someone points out to you 
that your pet theory of the universe is in disagreement with 
Maxwell's equations — then so much the worse for Maxwell's equations. 
If it is found to be contradicted by observation — well, these experimentalists do 
bungle things sometimes. But if your theory is found to be against the second law of 
thermodynamics I can give you no hope; there is nothing for 
it but to collapse in deepest humiliation.\rq\rq\ \cite{Eddington:1928}.
Quantum entanglement entropy cannot increase.}

There is another objection to this line of thought. Suppose the physics of the interior
of the black hole is controlled by same ideas of causality and local quantum field theory as the exterior.
Braunstein then showed that \lq\lq energetic curtains'' 
\cite{Braunstein:2009my,Braunstein:2011gz,Braunstein:2013mba,Braunstein:2014nwa} form on, or
just inside,  
the horizon. Subsequently, Almheiri, Marolf, Polchinski and Sully (AMPS) \cite{Almheiri:2012rt}
independently
discovered these ideas
and gave the name \lq\lq firewall'' to the same phenomenon.
For some later reviews of firewalls, 
see Harlow and Hayden \cite{Harlow:2013tf}, Almheiri, Marolf, Polchinski, 
Stanford and Sully \cite{Almheiri:2013hfa} and
Bousso \cite{Bousso:2013wia}.

Firewalls make the quandary just presented somewhat worse. Consider the totality of the Hawking radiation
that is described by some pure state. Divide the radiation up into its late and early components. At late
times the radiation consists of wavepackets with energy $\sim T_H$. Since asymptotic observers are believed
to be able to describe the evaporation process in terms of semi-classical physics, they can observe 
these radiated particles. At late times, the number of available states is small compared to early times
as the average energy of the emitted particles is increasing whereas the energy available in the
evaporating black hole is decreasing. The number of particles emitted is decreasing and  
we therefore expect the density matrix of the outgoing radiation to be getting closer 
and closer to describing 
a pure state. As a result, at late times 
one can construct creation and annihilation operators $b_i^\dag$ and
$b_i$ for these excitations. Furthermore, the number operator for the $i$-th state, 
$N_i=b_i^\dag b_i\ \   (no\ sum) $ will be (reasonably) well-defined because the states we are describing are 
very close to being pure. That means the entirety of the outgoing radiation
will be in an eigenstate of $N_i$.  Consequently, we can restrict our 
attention to the subspace of the Hawking radiation 
states for which this holds. These late outgoing states must be
 entangled with the early outgoing states which
themselves are the result of pair production outside the horizon. The ingoing early Hawking modes
can be described by creation and annihilation operators $a_j^\dag$ and $a_j$. The $a_j$ and $a_j^\dag$
are related to the $b_i$ and $b_i^\dag$ by Bogoliubov coefficients $A_{ij}$ and $B_{ij}$
in the way that was originally 
described in this context by Hawking \cite{Hawking:1974rv,Hawking:1974sw,Hawking:1976ra}.
Thus
\beq b_i  = \sum_j (A_{ij}a_j + B_{ij}a_j^\dag) \eeq
and therefore
\beq N_i = \sum_{jk} A^\ast_{ij}(A_{ik}a_j^\dag a_k + B_{ik}a_j^\dag a_k^\dag) + 
B_{ij}^\ast(A_{ik}a_j a_k + B_{ik}a_ja_k^\dag)  \eeq
Now suppose we are in the state $\vert S \rangle$ and look at the expectation value of $N_i$. 
We immediately see that if $\langle N_i \rangle \ne 0$, then $a_i\vert S \rangle \ne 0$.
Thus, we cannot be in the  $a$-vacuum. The firewall comes about because these modes will be 
Tolman blue-shifted relative to the asymptotic states. We conclude that the energy density blows up
close to the horizon.\footnote{In fact, this is almost apparent in Hawking's original treatment.}  
Whilst this treatment is approximate, it nevertheless poses an immediate problem for
our picture of black holes. 

A freely falling observer should be able 
to fall through the horizon without experiencing anything unusual. The principle of equivalence would be 
called into question were this not the case. If the firewall really exists, one of the central tenets of
relativity would  be violated.  Bousso has succinctly summarised 
the general reaction to firewalls, \lq\lq The firewall shakes the foundations of 
what most of us believed about black holes. It essentially pits quantum mechanics against 
general relativity, without giving us any clues as to which direction to go next,\rq\rq\
\cite{Bousso:oral}.

If unitarity of black hole evaporation is to be preserved, there must be something wrong with
our assumptions. The obvious place to look is to ask what happens in the interior of the black hole. 
We assumed that the fine-grained entropy was given entirely by the von Neumann entropy. 
However, the AdS-CFT correspondence has led to the conjecture that there is an extra contribution to the
fine-grained entropy. This idea originates with the observations of Ryu and Takayanagi 
\cite{Ryu:2006bv,Hubeny:2007xt,Lewkowycz:2013nqa,Faulkner:2013ana}. These ideas in some way parallel the
the Bekenstein-Hawking entropy of the horizon. The additional 
contribution to the fine-grained entropy $S_{RT}(X)$ in the context of black hole physics 
comes from a spacelike $2$-surface $X$ and is given by
$S_{RT}(X)=A(X)/4$ where $A(X)$ is the proper area of $X$. The definition of $X$, sometimes called the 
quantum extremal surface, is a little complicated, \cite{Akers:2019lzs,Almheiri:2019hni,
Engelhardt:2014gca}.
In the black hole context, 
take a partial Cauchy surface $\Sigma$ that has a single boundary anchored somewhere outside the 
black hole on an $S^2$ that completely encloses the horizon. $\Sigma$ extends into the interior of 
the black hole without either hitting the singularity
or having a further boundary. Now, if possible, find a minimal surface $Y$ on $\Sigma$. 
There may be no minimal surface on $\Sigma$ in which case $S_{RT}=0$
and indeed 
such a minimal surface will 
not generally exist outside the black hole. To find $X$, take $S(Y)$ with
\beq S(Y) =  \frac{A(Y)}{4} + S_{vN}(\Sigma_Y) \label{eq:RT} \eeq
with $S_{vN}(\Sigma_Y)$ being the von Neumann entropy of the quantum state 
on $\Sigma$ exterior to $Y$.  Now maximize $S(Y)$ by varying $\Sigma$ with its anchor 
fixed.  We might find multiple extrema by this process in which case it is the 
global minimum of (\ref{eq:RT}) that defines $X$. \cite{Almheiri:2020cfm}.
The new definition of the fine-grained entropy of a gravitating system is then
\beq S = S_{RT}(X) + S_{vN}(\Sigma_X). \eeq Ryu and Takayanagi interpreted $S_{RT}$ 
as being due to a CFT on the holographic screen provided by $X$. It is not clear what, if any, 
fundamental degrees of freedom describe $S_{RT}$ in the bulk.  

Because $X$ is determined by finding the global minimum of $S(Y)$, it seems that the definition of
$X$ should be extended by allowing for $\Sigma_X$ to contain disconnected components, 
\cite{Penington:2019npb,Almheiri:2019psf}. In this
case there could be \lq\lq islands\rq\rq\ of $\Sigma_X$ in the interior of $X$ as well as the 
exterior.  

These ideas have been used to calculate the entropy of the Hawking radiation and the entropy of 
the black hole interior and they reinforce the correctness of the Page curve \cite{Almheiri:2019hni}.
They do not 
give a way of reconstructing the quantum state of what gives rise to the black hole from the 
Hawking radiation. Nevertheless, they strongly suggest
that our attention should be directed to what happens
in the black hole interior. 

For a slightly different viewpoint that also suggests that physics in the 
black hole interior needs modification, see Akhoury, \cite{Akhoury:2013bia}.

\section{Consistent Histories}
\label{sec:coherent}

The idea of an observer causing wavefunction collapse seems rather alien to our present 
understanding of quantum mechanics. 
An observer 
is usually thought to be exterior to the system under observation. In the context of gravitational physics
this is not really possible except maybe for asymptotic observers in an asymptotically flat spacetimes.
More realistically, one should follow Everett \cite{Everett:1957hd} and Wheeler \cite{Wheeler:1957zz}
and think about participators and treat 
them as  quantum mechanical too. Such a reformulation of quantum mechanics has been achieved
(but perhaps not yet in its final form) by Griffiths \cite{Griffiths:1984rx,Griffiths:aa} and 
subsequently refined and expanded on by Omnes \cite{Omnes:1992ag} and by
Hartle and Gell-Mann \cite{GellMann:1992kh}.
There are excellent reviews by Hartle, \cite{Hartle:1992as, Hartle:2020fs}.

We assume that below the Planck scale we can neglect the quantum fluctuations in spacetime geometry
and take spacetime to be a differentiable manifold that has the usual causal structure and is time
orientable. Unless we are within a few Planck lengths of the singularity this seems to be a reasonable
assumption. We can take it that inside a black hole, but not too close to the singularity, 
conventional Hamiltonian evolution will occur and the usual rules of local quantum field theory apply.
The singularity is a place where the idea of spacetime as a smooth differentiable manifold
does not hold. We need to
keep an open mind about what takes place as one gets close to the singularity as it is there that
the established laws of physics must break down.

In the consistent histories approach to quantum mechanics, the basic concepts are a {\it state} and  
an {\it event}.
An event $E_i$ takes place at some specified time $t_i$ and  
its possible outcomes are labelled by $j$ taking the values $1$ or $0$ depending
on whether the result of the event $E_i$ is true or false.
$E_i$ is represented by a projection operator
on the Hilbert space of states. If the result of the event $E_i$ is $j$, then the projection operator
is $P_i^{(j)}(t_i)$. An event could  
have a discrete outcome, such as 
asking if a particle goes through a particular slit in a diffraction grating.
The outcome is either true or false depending on whether the particle went through the 
slit in question or not.
On the other hand  the event might probe part of a continuum, for example asking if a particle has an 
energy $E$ less
than $E_0$. Again there are only two possible outcomes for such an event. If the particle has energy 
less than $E_0$, the outcome is true, otherwise it is false.
In every case, each of the projection operators
can only have eigenvalue $0$ if false and $1$ if true. The projection operators have properties that 
reflect the propositional calculus of the $E_i$.
Summing over the possibilities $j$ for fixed $i$ must give unity
\beq \sum_j P^{(j)}_i = 1, \eeq
and the projections must be orthogonal 
\beq P^{(j)}_i P^{(k)}_i = \delta_{jk}P^{(j)}_i. \eeq
These are the quantum mechanical versions of  completeness
and exclusivity
in quantum logic \cite{BvN:aa, Mackey:aa}. Quantum logic is a set of rules that define the 
possible collection of events $E_i$. It is not the same as classical logic as it must prevent the 
occurrence of non-commuting projections at null separations.

An event can be many different things. It  could be an observation, the construction of a telescope
in some time interval, asking if the radioactive  decay of a particular nucleus has 
taken place or even if a coarse-graining of a more fundamental description has taken place.
In all cases it is a
restriction applied to the Hilbert space. However, the point of the consistent histories
approach is to include everything in a quantum mechanical setting by choosing a 
complete set of relevant projections. In doing so, we need to include the physics of the system 
being examined
together with the apparatus doing the examining and the observers. There is no longer any distinction 
between an quantum mechanical system and a classical observer as is required in the 
Copenhagen interpretation of 
quantum mechanics.

A history is a time ordered series of events. It is represented in the Hilbert space by a class 
operator $\C_\alpha$.
$\C_\alpha$ is a time-ordered series of projections, thus
\beq \C_\alpha = P_n(t_n)P_{n-1}(t_{n-1})  \ldots P_1(t_1) \eeq
with $t_n > t_{n-1} \ldots > t_1$.

Suppose one starts with a pure state in the Heisenberg picture $\vert \Psi \rangle$, 
then $\C_\alpha\vert\Psi\rangle$
is the branch of $\vert\Psi\rangle$ that corresponds to a history $\C_\alpha$. An alternative history 
might be $\C_{\alpha^\prime}$. If the state $\C_{\alpha^\prime}\vert\Psi\rangle$ is orthogonal to 
$\C_\alpha\vert\Psi\rangle$ then the two states are said to be consistent. One can think of
this as saying that such pairs of states are independent of each other. Suppose there is a collection of
histories $\{\C_\alpha\}$,  one can then form the decoherence functional $D(\alpha,\alpha^\prime)$
for all elements  $\alpha,\alpha^\prime \in \{\alpha\}$.
\beq D(\alpha,\alpha^\prime ) = 
\langle\Psi\vert \C^\dag_\alpha \C_{\alpha^\prime}\vert\Psi\rangle .\label{eq:decohere} \eeq 
If $D(\alpha,\alpha^\prime)$ is diagonal, 
then whole set of histories $\{\C_{\alpha}\}$ is said to decohere or be consistent.
Under these circumstances, the probability $p_\alpha$ of the history $\C_\alpha$ 
is given by
\beq p_\alpha = \langle\Psi\vert \C^\dag_\alpha \C_\alpha\vert\Psi\rangle. \eeq
It might be that $D(\alpha,\alpha^\prime)=0$ with $\alpha\ne\alpha^\prime$ 
does not hold exactly, but is subject to small violations. 
Under these circumstances it is still possible to find probabilities but they will be subject to
some kind of approximation. This type of behaviour could occur if some coarse-graining were
not sufficiently precise.

It is straightforward to extend the computation of probabilities from pure states $\vert\Psi\rangle$
to  states described by a density matrix
$\rho_i$,
\beq p_\alpha = tr \langle \C^\dag_\alpha\rho_i\C_\alpha\rangle. \eeq	 
Now we need to see if we can rewrite this expression as a path integral. Suppose that we choose 
a series of spacelike surfaces labelled by a time co-ordinate and projectors, or a coarse-graining, 
that selects out the metric and other quantum fields everywhere on these surfaces. Then we can construct 
a path integral, as explained in \cite{Hartle:1992as}. By doing this, we are ignoring the 
fundamental nature of quantum gravity as we do not have much of an idea of what it is. This is a particular
kind of coarse-graining. One can then hope 
that the decoherence functional is well-behaved. The result is a path integral version of
$D(\alpha^\prime,\alpha)$ that is essentially the same as the 
Schwinger-Keldysh \cite{Schwinger:1960qe,Baym:xx,Keldysh:1964ud}  path integral
and in which 
\beq D(\phi_i,\phi_i^\prime) = \int D[\phi] D[\phi^\prime] \delta(\phi_f-\phi^\prime_f) \rho_i(\phi_i,\phi^\prime_i) e^{i(I[\phi]-I[\phi^\prime])}.
 \eeq
$\phi$ represents all of the fields in the problem and $I[\phi]$ is the action.  
The path integral in the forward direction is taken over all fields that start with $\phi_i$ 
at time 
$t_i$ and end with $\phi_f$ at time $t_f$. The path integral over the backward direction 
is taken over
fields that end at time $t_f$ with the field being $\phi_f^\prime$ and start at time $t_i$ with the field
being $\phi_i^\prime$ in a way that is consistent with the density matrix 
describing the initial state being
$\rho_i(\phi_i,\phi_i^\prime)$. Since it is the initial state being specified by some density matrix, there
is no requirement that $\phi_i=\phi_i^\prime$
For the final state, the delta function enforces the matching condition that
$\phi$ and $\phi^\prime$ be the same. 
 
\section{Time Symmetric Quantum Theory}
We are familiar with the idea that the future is different to the past.
There are many arrows of time
that show us in what way the future differs from the past \cite{Penrose:xy}. 
Despite this, the fundamental laws of physics are time reversal invariant.

Perhaps the most familiar 
is psychological time,
in that we remember our own past whilst the future remains a mystery until we experience it. Each of us 
is what Gell-Mann calls an IGUS (Information Gathering and Utilizing System) \cite{Gell-Mann:2018dzd}, 
that is an entity capable
of making observations and drawing conclusions from them. However, there is also some ambiguity about 
psychological time because of relativistic effects. Different observers may see the same events 
but in various orders and at different times. That is because each IGUS is an independent entity.
That makes psychological time a concept that has a strong subjective component. Nevertheless, an 
observation
can correspond to a quantum mechanical event and result in the insertion of a projection operator  into the 
decoherence functional.  

There are two arrows of time that appear unambiguously in classical physics. The first is the observation
in electrodynamics that the fields due to some distribution of currents and charges is governed by
the retarded solution to the inhomogeneous wave equation. In effect this is the observation that cause
precedes effect and of course applies not only to electrodynamics but much more widely. 
Sound waves travel from the 
source to a receiver; not the other way around. Waves spread out from where a stone is thrown into a pond.
The second is the observation that the Universe is expanding. That is inferred from the fact the galaxies
exhibit a red-shift that grows with distance and also from the existence of 
the cosmic microwave background 
that shows that the universe has evolved from a very dense early state. It appears that the Universe 
originated in some kind of initial singularity as predicted by the singularity theorems of Hawking 
and Penrose \cite{Hawking:1969sw}. As an aside, it presently appears that the Universe will 
never recollapse to a 
big crunch singularity but will expand forever powered by a small positive cosmological constant.

The second law of thermodynamics, that the entropy of an isolated system cannot decrease, defines a
thermodynamic arrow of time. However, this arrow of time is really about complicated systems 
where there has been some kind of coarse-graining. A system that has some degree of order will evolve 
in a way given 
by the microscopic laws specifying the evolution of its most fundamental components. However, if you
only look at the bulk properties of the system, most of the time it will appear to have become 
more disordered
reflecting the randomizing nature of the coarse-graining process. That the second law is of limited
applicability is demonstrated by the phenomenon of Poincar\'e recurrence. 

In quantum mechanics, at first sight, there does seem to be a time direction. In the 
Copenhagen interpretation, where an observation is taken to result in the collapse of
the wavefunction, there is clearly a time asymmetry.  However, in post-Everett quantum mechanics
as exemplified by decoherent histories,
there is no collapse of the wavefunction, merely an application  of a projection onto the
state of the system. Of course, as we saw in section \ref{sec:coherent}, projections are 
required to be time ordered, but appear in the decoherence functional in a time symmetric way.
That is to say the time-ordered history $\C_{\alpha}$ appears together with the anti-time-ordered 
history $\C^\dag_{\alpha}$. Additionally, there is no analogue of the second law of thermodynamics, 
since in a closed system the von Neumann entropy is constant. 

There is one remaining type of time asymmetry and that is provided by the CP-violation in $K_0$
decay. It is strongly believed that CPT is an exact symmetry of nature and so CP-violation 
translates into a T-violation. However, this is a very weak phenomenon and can be accounted for  
in the standard model. It appears unrelated to the other arrows of time. 

As just observed, CPT is believed to be an exact symmetry of nature. Nevertheless, arrows of time
are real and must originate from somewhere. The action of CPT on a particle is to transform   
the particle into its antiparticle, perform a total spatial reflection and reverse its momentum. 
Once this has been done, CPT symmetry says that the same laws of physics apply now as 
to the original system. How then can there be any arrow of time? The answer can only come from the 
boundary conditions. Long ago, Einstein, Tolman and Podolsky \cite{Einstein:1931zz} showed that 
in quantum mechanics once a measurement is made,  not only does this result in a probabilistic view 
of the future but also it means you have a similar probabilistic view of the past. Thus quantum mechanics 
does not have an inbuilt arrow of time.

It therefore seems strange that the decoherence functional contains only a contribution from the initial 
density matrix but makes no reference to the future. Were one trying to retrodict the past rather than 
predict the future, the decoherence functional would contain a density matrix at 
its future endpoint instead.  This observation stimulated Aharonov, Bergmann and Lebowitz 
\cite{Aharonov:xz}
to suggest that under certain circumstances one should include in the decoherence functional both
an initial state density matrix and a final state density matrix.
The decoherence functional is then replaced by
\beq D(\alpha,\alpha^\prime) = \frac {tr\  (\rho_f\C^{\dag}_{\alpha}\rho_i\C_{\alpha^\prime})}
{tr\ (\rho_f\rho_i)} \eeq
There is the trivial possibility that $\rho_f$ is the identity. This is what Gell-Mann and Hartle
\cite{Gell-Mann:2018dzd} 
call the principle of indifference and under these circumstances the conventional presentation 
of consistent histories is 
regained. For a more detailed discussion of the decoherence functional in time symmetric quantum
mechanics, see Hartle, \cite {Hartle:2020his}.

What we would like to investigate now is the possibility that the principle of indifference 
holds outside the black hole, but as one approaches the singularity, or more precisely what classically
is called the singularity, one needs to specify some particular $\rho_f$. 

One might feel sceptical that a boundary condition in the future makes sense. However,
in certain circumstances open
to experimental verification, \cite{Aharonov:az,Leggett:aa,Lloyd:2010nt}
these ideas have proved to be correct. 

\section{Singularities}
Our ideas about black hole information loss are conditioned by the classical picture of black holes in 
general relativity. The singularity theorems show that once there is an horizon, the spacetime is singular.
Some intuition can be derived from spherical collapse where a spacelike singularity forms which is the 
future boundary of all causal lines that get trapped inside the horizon.  Classically, 
the generic situation appears to be that collapse results in a 
spacelike, or possibly a null,  singularity in the future of all worldlines that pass through the horizon. 
Such worldlines terminate at the singularity. The singularity is the future boundary 
of the part of spacetime inside the horizon and so anything reaching it can no longer be thought of as
being in the spacetime. 

We need to study the nature of this singularity. Let us first examine what happens in classical
 physics. The most convenient approach for present purposes is to use the canonical formalism,
 \cite{Dirac:1958sc,Arnowitt:aa,DeWitt:1967yk,Hanson:1976cn}.
Spacetime is taken to be foliated by spacelike surfaces $\Sigma(t)$ with spatial coordinates $x^i$,
a time coordinate $t$.  The induced metric on $\Sigma(t)$ is $\gamma_{ij}$. The line element 
for the spacetime
can then by written as
\beq
ds^2 = -N^2dt^2 + \gamma_{ij}(dx^i+N^idt)(dx^j+N^jdt)
\eeq
where $N$ is the lapse and $N^i$ the shift. $N, N^i$ and $\gamma_{ij}$ are in general functions of both
$x^i$ and $t$.  In what follows we will deal with pure Einstein gravity where the action is, including the
boundary term \cite{York:1972sj,Gibbons:1976ue},
\beq
I = \int_{\cal M} R\ (-g)^{1/2}\ d^4x \pm 
 2\ \int_{\partial{\cal M}} \nabla_an^a\ (\pm\sigma)^{1/2}\ d^3x 
\eeq
with ${\cal M}$ being the spacetime manifold, ${\partial{\cal M}}$ its boundary, $R$ the Ricci scalar
of the four-dimensional spacetime and $n^a$ the unit normal to the boundary
where the induced metric is $\sigma$ and the signs are determined by whether the boundary is 
spacelike or timelike. Using the $3+1$ decomposition of the metric, the action becomes
\beq I = \int_{\cal M} (K_{ij}K^{ij}-K^2 + {}^{(3)}R(\gamma))\ N\ \gamma^{1/2} \ d^3x\ dt \eeq
where ${}^{(3)}R(\gamma)$ is the Ricci scalar of $\gamma_{ij}$,  $K_{ij}$ is the second fundamental
form describing the embedding of $\Sigma(t)$ in ${\cal M}$ and $K=\gamma^{ij}K_{ij}$. Explicitly
\beq K_{ij} = \frac{1}{2N}\Bigl(D_iN_j +  D_jN_i - \frac{\partial \gamma_{ij}}{\partial t}\Bigr) \eeq
with $D_i$ being the covariant derivative with respect to the three-metric $\gamma_{ij}$. 

The momentum conjugate to $\gamma_{ij}$ is $\pi^{ij}$ given by
\beq
\pi^{ij} = -\gamma^{1/2}\Bigl(K^{ij}-\gamma^{ij}K\Bigr).
\eeq
The momenta conjugate to both $N$ and $N^i$ vanish and this results in a set of constraints.
The diffeomorphism constraint coming from the vanishing of the momentum conjugate to $N^i$ is
\beq \chi^i \equiv D_j\pi^{ij} = 0,  \eeq
If it is satisfied at one instant of time, since it does not evolve, it will always be satisfied
and so is a constraint on the data on $\Sigma$ at any moment. The other constraint is
the Hamiltonian constraint
\beq {\cal H} \equiv \Bigl({\cal G}_{ijkl}\pi^{ij}\pi^{kl} - \gamma^{1/2}\ {}^{(3)}R\Bigr) =0.
\eeq
with
\beq
{\cal G}_{ijkl} = \frac{1}{2}\gamma^{-{1/2}}(\gamma_{ik}\gamma_{jl} + \gamma_{il}\gamma_{jk} - 
\gamma_{ij}\gamma_{kl}). \eeq
The Hamiltonian constraint bears a superficial resemblance to the mass-shell condition
for a massless particle, which suggests that one should interpret ${\cal G}_{ijkl}$ as the inverse metric
on the space ${\bf M}$ of all metrics $\gamma_{ij}$ at each point in space. The metric ${\cal G}^{ijkl}$
on ${\bf M}$ can then calculated from 
\beq {\cal G}^{ijkl}{\cal G}_{klmn} = \frac{1}{2}(\delta^i_m\delta^j_n + \delta^i_n\delta^j_m) \eeq
and is 
\beq {\cal G}^{ijkl} = \frac{1}{2}\gamma^{1/2}(\gamma^{ik}\gamma^{jl}+\gamma^{il}\gamma^{jk}
-2\gamma^{ij}\gamma^{kl}). \eeq.

${\cal G}^{ijkl}$ has signature $(-++++\, +)$ with the negative direction being associated with 
conformal transformations. It should be carefully noted that the negative direction has 
nothing whatsoever to with the time direction in the original spacetime. In fact, the negative
direction associated with conformal transformations is the same as that making the Euclidean formulation
of gravity somewhat sick \footnote{Historically, the negative direction has been 
termed \lq\lq extrinsic time\rq\rq\ which is grossly misleading.} \cite{Gibbons:1978ac}.  
The conformal factor can vanish and when this happens  ${\bf M}$ is singular. 
DeWitt \cite{DeWitt:1967yk} refers to the singularity in ${\bf M}$ as the \lq\lq frontier\rq\rq\ and it is 
an avatar of the spacetime
singularities we are ultimately interested in. Quotienting out the conformal degree of freedom results in 
${\cal G}^{ijkl}$ being the metric on the symmetric space  $SL(3,{\mathbb R})/SO(3)$.
Finally, in order to eliminate gauge degrees of 
freedom conjugate to the constraints, one must fix both the lapse and shift and 
eliminate the diffeomorphisms from the metric $\gamma_{ij}$.

Although we expect to encounter a singularity inside a black hole, the picture presented by
Oppenheimer and Snyder \cite{Oppenheimer:1939ue} is somewhat misleading. 
The reason is that one does not expect exact 
spherical symmetry. The general approach to a singularity appears to be chaotic as was first shown by
BKL, \cite{Belinsky:1970ew,Belinsky:1982pk,Belinsky:aa}.  
Their results indicate that as one gets close to the singularity, each point of
the spacetime behaves more or less independently of neighbouring points. That is not to say 
that neighbouring points do not interact,
but rather that as one gets close to the singularity, their influence can be summarised in a way 
that results in a considerable simplification of the Einstein equations. Such a behaviour is termed
ultralocal. The results of BKL show that
the time evolution at each point can be described in terms of ordinary differential equations for 
metric components, with the independent variable being time. 
In the BKL idealisation, each point
is independent but in reality it  seems that by point, one really means an appropriately small spatial 
region.  Scheel and Thorne \cite{Scheel:2014hfa} 
suggested that perhaps the BKL picture of singularity formation was relevant to black hole physics. 
Recent numerical work \cite{Garfinkle:2020lhb} is consistent with the BKL picture of the 
approach to singularities.

The starting point for an exploration of the BKL approximation is a minisuperspace \cite{Misner:1972js} 
description of the Kasner universe.
One firstly partially fixes the gauge by choosing pseudo-Gaussian normal coordinates for a Bianchi I 
spacetime. The metric is thus
\beq ds^2 = -N^2 dt^2 + a(t)^2dx^2 + b(t)^2dy^2 + c(t)^2dz^2
\eeq
with the lapse being a function only of $t$. 
Substituting the above metric form into the Einstein action gives 
\beq I = \int dt \ \Biggl[-\frac{2}{N}\Biggl(\dot a \dot b c + \dot b \dot c a + \dot a \dot c b\Biggr)
\Biggr] \label{eq:kasneract}\eeq
where a dot denotes the derivative with respect to $t$.
The solutions are given by
\beq a=a_0(t_0-t)^{p_1},\ \ \ \ b=b_0(t_0-t)^{p_2},\ \ \ \ c=c_0(t_0-t)^{p_3} \label{eq:kasner} \eeq
with the Kasner exponents $p_i$ being given by
\beq  p_1 + p_2 + p_3 = 1, \ \ \ \  p_1^2 + p_2^2 + p_3^2 = 1 \eeq
and $a_0, b_0$, $c_0$ and $t_0$ being constants.
Thus, as is well-known, two of the $p_i$ are positive and one negative.  
Hence as $t$ approaches $t_0$ from below, two directions are contracting and one is expanding \footnote{
The Kasner universe is usually presented as starting at a singularity and then two directions
are expanding. Here we are looking at a collapsing version of the Kasner universe so two directions are 
contracting and one is expanding.}.
At $t_0$ there is a curvature singularity although the Ricci scalar remains zero. 
The volume of space, which is proportional to the product $abc \sim (t-t_0)$, tends to zero as
 $t \rightarrow t_0$ from below.
It should be noted that when the action is varied with respect to $N$, it generates the Hamiltonian
constraint. 

A more convenient set of variables are
\beq a=e^{-\beta^1},\ \ \ \ b=e^{-\beta^2},\ \ \ \ c=e^{-\beta^3} \eeq
and 
\beq \tilde N = \frac{N}{abc} \eeq
in terms of which the action becomes
\beq I =
 \int dt \Biggl[-\frac{2}{\tilde N}\Biggl(+\dot\beta^1\dot\beta^2+\dot\beta^2\dot\beta^3+\dot\beta^3\dot\beta^1\Biggr) \Biggr].\eeq
The momenta $\pi_i$ conjugate to $\beta^i$ are
\beq \pi_i = \frac{2}{\tilde N} {\cal G}_{ij}\dot\beta^j \eeq
where the inverse minisuperspace metric is
\beq {\cal G}_{ij} = \begin{pmatrix} 0&\ \ -1&\ \ -1\\-1&\ \ 0 &\ \ -1\\-1&\ \ -1&\ \ 0 \end{pmatrix}. \eeq
The Hamiltonian constraint can now be written as 
\beq {\cal H} = \frac{1}{4\tilde N}{\cal G}^{ij}\pi_i\pi_j \eeq
with the minisuperspace metric being
\beq {\cal G}^{ij} =\frac{1}{2}\ \begin{pmatrix} 1&\ \ -1&\ \ -1\\-1&\ \ 1&\ \ -1\\-1&\ \ -1&\ \ 1 \end{pmatrix}. \eeq
${\cal G}^{ij}$ has signature $(-++)$, is flat and the negative direction is the overall scale. 
Solutions to the Einstein equations can be seen to be null geodesics in this space. 
The affine parameter $\tau$ on these geodesics is a measure of $t$ in which 
$d\tau=\tilde Ndt$. These geodesics are of the form
\beq \beta^i = w^i\tau + \beta_0^i \eeq
with both $w^i$ and $ \beta_0^i$ being constants. $w^i$ 
is null in the sense that ${\cal G}_{ij}w^iw^j=0$. The $w^i$ are related to the Kasner exponents $p^i$
defined in (\ref{eq:kasner}) by
\beq p_i = \frac{w^i}{w^1+w^2+w^3}.\eeq
Defining $\rho$ by
\beq \rho^2 = -\beta^i\beta^j{\cal G}_{ij} \eeq
one observes that the singularity where $t=t_0$ is approached as $\rho \rightarrow \infty$
where the proper volume of space tends to zero. 

A more geometrically pleasing understanding of the minisuperspace picture introduced above 
comes from exhibiting its
relationship to the Lobachevskii plane, \cite{Belinsky:aa,Damour:2002et} the two-manifold of 
constant negative curvature,
with polar coordinates $0\le r<\infty,\phi\le 0\le 2\pi$.
Let $\beta^i=\rho \gamma^i$
then
\begin{align} &\gamma^1 = 
\frac{1}{\sqrt{6}} \cosh r - \sqrt{\frac{2}{3}}\sin\Biggl(\phi+\frac{\pi}{3}\Biggr)\sinh r \\
 &\gamma^2 = \frac{1}{\sqrt{6}} \cosh r - \sqrt{\frac{2}{3}}\sin\Biggl(\phi-\frac{\pi}{3}\Biggr)\sinh r \\
 &\gamma^3 = \frac{1}{\sqrt{6}} \cosh r + \sqrt{\frac{2}{3}}\sin\phi \sinh r \end{align}
The minisuperspace line element is now
\beq d\Sigma^2 = -d\rho^2+\rho^2(dr^2 + \sinh^2r d\phi^2) \eeq

The BKL approximation is a relatively simple modification of this picture. The influence of 
neighbouring cells can be described by the introduction of perfectly reflecting walls 
constraining the null geodesic motion to the spatial region 
\beq \gamma^1-\gamma^2 \ge 0,\ \ \ \ \gamma^3-\gamma^2 \ge 0, \ \ \ \ \gamma^1 \ge 0 \eeq
or equivalently in the Lobachevskii plane 
\beq \frac{\pi}{6}\le \phi \le \frac{\pi}{2}, \ \ \ \ \coth r \ge 2\sin(\phi+\frac{\pi}{3}). \eeq
At these walls, the geodesics are specularly reflected. Back in the Kasner picture, the 
effect of a reflection is to change the Kasner exponents and rotate the principal axes of the
Kasner expansion or contraction. 

Remarkably, mapping the spatial sections into the upper-half Poincar\'e plane reveals
some hidden structure. Let
\beq x^0=\cosh r,\ \ \ \ x^1=\sinh r\sin\phi,\ \ \ \ x^2=\sinh r\cos\phi. \eeq
Now set
\begin{align} &\hat x^0 = \frac{2}{\sqrt{3}}x^0 - \frac{1}{\sqrt{12}}x^1 - \frac{1}{2}x^2, \\
&\hat x^1 = \frac{1}{\sqrt{3}}x^0 - \frac{1}{\sqrt{3}}x^1 - x^2, \\
&\hat x^2 = \frac{\sqrt{3}}{2}x^1 - \frac{1}{2}x^2, \end{align}
and then transform the \lq\lq spatial\rq\rq\ part into the upper-half complex plane
using
\beq z = u+iv = \frac{\hat x^1 + i(\hat x^0+\hat x^2+1)}{\hat x^0-\hat x^2+1+i\hat x^1}. \eeq
The classical behavior of this system is that of null geodesics in the metric
\beq ds^2 = -d\rho^2 + \rho^2\, \Bigl(\frac{du^2 + dv^2}{v^2}\Bigr). \label{eq:ssm} \eeq
The action in terms of $\rho,u$ and $v$ can now be taken to be
\beq I = \int dt\ \  \frac{1}{2}\Bigl[ \ -\dot\rho^2 + \frac{\rho^2}{v^2}(\dot u^2 + \dot v^2)\Bigr]. \eeq
The momenta are defined by
\begin{align} &\pi_\rho  = -\dot\rho \\ &\pi_u = \frac{\rho^2}{v^2}\dot u\\ 
&\pi_v = \frac{\rho^2}{v^2}\dot v \end{align}
The Hamiltonian is now 
\beq {\cal H} = -\frac{1}{2}\pi_\rho^2 + \frac{v^2}{2\rho^2}(\pi_u^2 + \pi_v^2) \eeq
which must be supplemented by the constraint that ${\cal H}$ must vanish.
The walls are now located at 
\begin{align}  &{\text{Wall}}\ 1:\ \  u=0,\ \  v \ge 1\\
 &{\text{Wall}}\ 2: \ \ u=\frac{1}{2},\ \  v \ge \frac{\sqrt{3}}{2} \\
 &{\text{Wall}}\ 3: \ \ z=e^{i\theta},\ \ 
\frac{\pi}{3} \le \theta \le \frac{\pi}{2}. \end{align}
This is a fundamental region $F$ of $PGL(2,\mathbb Z)$ and is precisely half of the more familiar 
fundamental region of $PSL(2,\mathbb Z)$, see Figure $2$. 
Again, there is specular reflection from the walls
and this has the effect of transforming the trajectory, $z(\rho)$ close to the wall
into $R_i(z)$ as
\begin{align} &{\text{Wall}}\ 1:\ \ R_1(z) = -z^\ast \\
&{\text{Wall}}\ 2:\ \ R_2(z) = 1-z^\ast \\
&{\text{Wall}}\ 3:\ \ R_3(z) = \frac{1}{z^\ast}.\end{align}
A series of reflections then is described by a word in this group. The Kasner evolution 
classically will typically involve an infinite number of reflections before reaching the singularity 
at $\rho \rightarrow \infty$. It should be noted that the generators of $PSL(2,\mathbb Z)$
are words in $PGL(2,\mathbb Z)$.  
The generator of translations in $PSL(2,\mathbb Z)$ is $T:z\rightarrow z+1$ and therefore $T=R_2R_1$.
Similarly $S:z \rightarrow -\frac{1}{z}$ is $S=R_1R_3$. Although it is not of obvious 
direct relevance here, 
$PGL(2,\mathbb Z)$ is the Weyl group of the Kac-Moody algebra $A_1^{++}$. The Weyl group describes 
how the root vectors can be chosen and this in turn determines where we decide to place the 
fundamental region.  

To summarise:   classically, there are domains that for some period look like the Kasner universe
in the sense that two directions are contracting and one is expanding.
At the beginning and end of each period there is a discontinuity and the exponents and the directions 
of expansion and contraction change as a 
result of bouncing off the walls. To reach the singularity, there are generically an infinite number
of bounces. However, the volume of any cell goes to zero as the singularity is approached since 
the product $abc$ is  proportional to $(t_0-t)$. This type of behaviour is what is expected classically 
as one approaches the singularity in any realistic collapse.

\section{Quantum Gravity}

How is the situation different in a quantum universe as compared to a classical universe?
For quantum field theory in curved spacetimes, the situation seems to be  essentially 
similar to the classical one. 

Consider the Schwarzschild spacetime and massless scalar fields propagating 
in it, \cite{Futterman:1988ni,Castro:2013lba}. The radial part of 
the Klein-Gordon equation has a resonant singularity at $r=0$ and the solutions of the Klein-Gordon
equation have logarithmic singularities there. As a result, the probability flux vector for 
excitations crossing the horizon is future-pointing there and divergent and there  
appears to be an inevitable loss of information. Similar results hold for fields with spin or mass.
The Kerr spacetime is just as bad, since the inner horizon is unstable for essentially the same reason, 
\cite{Simpson:1973ua}.
Solutions of the radial part  of the Teukolsky equation blow up on the inner horizon and it is for this 
reason that one believes a singularity to be formed there.

To make any progress, one should be looking at quantum gravity. There is no fine-grained 
theory of quantum gravity known. The best we can do is to look at a coarse-grained approach based on
general relativity. The classical theory is the most coarse-grained picture. Semi-classical
quantum gravity based on general relativity is less coarse-grained and
allows for some exploration of the quantum world.
One can explore semi-classical gravity either by the use of path integrals or via the 
canonical formulation. In what follows, we will motivate the use of the Wheeler-DeWitt equation
by observing that the wavefunction of the universe derived from the path integral is a solution of
the Wheeler-DeWitt equation. 

Path integral expressions for a wavefunction in gravitation are of the form
\beq \Psi[\gamma] = \int D[g] e^{iI[g]} \eeq
where $I[g]$ is the classical action and the integral is taken over all distinct metrics subject to
the metric on the boundary being $\gamma$. $\Psi$ will obey the Wheeler-DeWitt equation.
The Hamiltonian constraint ${\cal H}$ is in general a functional of components of the 
metric $\gamma_{ij}$ and 
the canonical
momenta $\pi^{ij}$. To quantize, we replace the momenta  $\pi^{ij}$ by $-i{\delta\over\delta\gamma_{ij}}$.
Then the Wheeler-DeWitt equation is
\beq {\cal H}\Psi = 0  \label{eq:wdw} \eeq
and in some sense replaces the Schr\" odinger equation for gravity.  One must also 
impose the diffeomorphism
constraint so that in addition
\beq \chi^i \Psi = 0. \label{eq:diff} \eeq
Of course, the prescription for
determining the wavefunction needs to be supplemented by physically appropriate boundary conditions.

Hartle and Hawking {\cite{Hartle:1983ai} defined the scalar product of two wavefunctions $\Psi_a$ and
$\Psi_b$  to be
\beq 
(\Psi_a, \Psi_b)_{HH}  =  \int D[\gamma]\ \Psi_a^\ast[\gamma]\ \Psi_b[\gamma]. \label{eq:herm} \eeq
The Hartle-Hawking inner product is the integral over all spatial metrics at each point in space.
We immediately recognise (\ref{eq:herm}) as being related to the decoherence functional for 
gravity,  (\ref{eq:decohere}). 

An alternative definition of the inner product is due to DeWitt,
\beq (\Psi_a, \Psi_b)_{DW} = \int\ D[\bar\gamma]_{ij}\ {\rm Im} 
\Biggl(\Psi_a[\gamma]^\ast\ {\cal G}^{ij}{}_{kl}
 \frac{\delta}{\delta\gamma_{kl}}\Psi_b[\gamma]\Biggr). \label{eq:dwnorm} \eeq
 In this expression, the integral is again evaluated at each point in space, but instead of 
 integrating over all metrics $\gamma$, one integrates over all metrics $\bar\gamma$ that have 
 the conformal factor quotiented out, \cite{DeWitt:1967yk}.
 At each point in space, one integrates over the five-dimensional collection of metrics $\bar\gamma_{ij}$
 that parametrise the symmetric space $SL(3,\mathbb R)/SO(3)$. The subscript on the measure
 $D[\bar\gamma]_{ij}$ indicates that this is a vector in the space of all metrics in the direction of
 conformal transformations. In many ways, this is a functional analog of the Klein-Gordon norm
 where one is integrating the divergence-free Klein-Gordon current over a spacelike surface.
 Here one is integrating at each 
 point in space a vector, the DeWitt current, that is similarly divergence-free in a functional way as a 
 result of obeying the Wheeler-DeWitt equation. 
 The surface one is integrating over is \lq\lq spacelike\rq\rq\ as it is the conformal direction in the 
 space of all metrics that has a \lq\lq timelike\rq\rq\ direction. 
 Because of the divergence-free condition,\ (\ref{eq:dwnorm})\ is independent of precisely how one 
 chooses the 
 hypersurface determining $\bar\gamma \subset \gamma$ as long as the hypersurface intersects each
 conformal equivalence class once only.
   
Consider now the Wheeler-DeWitt equation for our minisuperspace describing the approach to 
the singularity. It is just the vanishing of the  Laplacian of the metric (\ref{eq:ssm}) acting on 
$\Psi$. However, we need to take account of the effect of the walls. The walls generate specular 
reflection classically and therefore mark the locations where the potential becomes infinite
and thereby restricts classical motion to the fundamental region, $F$. For the Wheeler-DeWitt 
equation this is equivalent to introducing a potential that is infinite beyond the walls. 
Standard arguments 
about the solution to differential equations of this type tell us that $\Psi$ must vanish on the walls
that make up the boundary of the fundamental region. 
 
 For our minisuperspace the Wheeler-DeWitt equation becomes
 \beq \rho^2\frac{\partial^2\Psi}{\partial\rho^2} + 2\rho\frac{\partial\Psi}{\partial\rho}
 +\Delta_F \Psi = 0 \eeq
 where $\Delta_F$ is the Laplacian in the fundamental region. 
 An explicit expression for $\Delta_F$ is 
 \beq \Delta_F = -v^2\Bigl(\frac{\partial^2}{\partial u^2} + \frac{\partial^2}{\partial v^2}\Bigr). \eeq
 One could solve the Wheeler-DeWitt equation in terms of Maass functions and appropriate cusp forms for
 $F$, \cite{Terras:aa}. Fortunately,  a detailed computation of $\Psi$ is unnecessary here.  The  eigenfunctions $f_n$ of $\Delta_F$ obey
 \beq \Delta_Ff_n = s_n(1-s_n)f_n. \eeq
 The $f_n$ are the odd Maass waveforms of $SL(2,\mathbb Z)$ with $s_n=\tfrac{1}{2}\pm it_n$
 with $t_n$ real. For each $n$, there is a complex conjugate pair of such eigenfunctions.
 There are two families of such wavefunctions. The first is a discrete set of odd cusp forms of 
 $SL(2,\mathbb Z)$ that are square integrable in $F$ with the norm inherited from constant negative
 curvature space,
 \beq \int_F \frac{dudv}{v^2} \ f_n^\ast f_n. \eeq
 The second is 
 a continuum of odd non-holomorphic Eisenstein series
 that are not integrable in this norm. 
 Despite the fact that the non-holomorphic Eisenstein series are not normalizable, 
 normalizable functions obeying our boundary conditions are represented by linear 
 combinations of members of both families.
 
 Suppose now that solutions of the Wheeler-DeWitt equation are separable and can be written in the form 
 $\sum _n P_n(\rho) f_n(u,v)$. The $P_n(\rho)$ obey
 \beq \frac{\partial^2P_n}{\partial\rho^2}+\frac{2}{\rho}\frac{\partial P_n}{\partial \rho}
 +\frac{\lambda_nP_n}{\rho^2} = 0. \eeq
 Solutions to the \lq\lq radial\rq\rq\ part of the Wheeler-DeWitt equation are then just powers
 $ P_n(\rho) = \rho^{\, p_n}$ with $p_n = -\tfrac{1}{2} \pm it_n$. Thus as 
 $\rho\rightarrow\infty$, $\Psi\rightarrow 0$.

The two inner products $(\Psi_a,\Psi_b)_{HH}$ and $(\Psi_a,\Psi_b)_{DW}$ both descend to 
inner products on our minisuperspace. Explicitly,
\beq 
(\Psi_a,\Psi_b)_{HH} = \int \ \Psi_a^\ast\, \Psi_b\ \ \frac{\rho^2}{v^2}\  du\, dv\eeq
and
\beq
(\Psi_a,\Psi_b)_{DW} = \int \ {\rm Im} \  \Bigl(\Psi_a^\ast \
\frac{\partial}{\partial\rho}\ \Psi_b \Bigr) \ \frac{\rho^2}{v^2}\ d\rho\, du\, dv. \eeq
Assuming that $\Psi$ is derived from functions that are normalizable on $F$, they are 
normalizable in the DeWitt norm, but not in the Hartle-Hawking norm. 

From the path integral form of $\Psi[\gamma]$, one notes that $\Psi[\gamma]$ is the probability amplitude
for finding the geometry described by the spatial metric $\gamma$. In the BKL minisuperspace, the
geometry is described by $\rho,u$ and $v$ or equivalently $a,b$ and $c$. So imagine asking 
for the probability amplitude of finding the geometries with $\rho\rightarrow\infty$ corresponding 
to surfaces getting closer and closer to what classically is singular. For each eigenfunction $f_n$ 
of $\Delta_F$, 
\beq \Psi \sim \rho^{-{1/2}} e^{\pm it_n\ln \rho} f_n. \label{eq:wvfn} \eeq
The probability of getting close to the singular surface is decreasing to zero. 
Presumably there
is no singularity \footnote{ 
DeWitt \cite{DeWitt:1967yk} speculated that $\Psi$ would 
have to be set to zero at the frontier of $M$. }.
There is however an
oscillatory part of this wavefunction that looks like waves travelling both forward and backwards in
$\rho$ depending on the sign chosen. For a given $f_n$, the choice of sign is completely arbitrary
so one can always choose $\Psi$ to contain arbitrary linear combinations of both possible signs.
In many ways, the DeWitt current looks like a probability current so if information is not to be lost
close to the singularity, one wants not just for $\Psi$ to vanish, but the DeWitt norm also to vanish.
Since each eigenfunction is of order $\rho^{-{1/2}}$ one concludes that the DeWitt norm generically 
independent of $\rho$. But it is easy to make the DeWitt norm vanish by making $\Psi$ real as can be
done by replacing $\Psi$ by $\Psi+\Psi^\ast$, as this will still obey the Wheeler-DeWitt equation. 
This last condition prevents information from escaping from what appears to be the future boundary 
of the spacetime.

We propose to take seriously this boundary condition on the surfaces close to the classical singularity.
It amounts to setting a boundary condition in the future that enforces a reflection in the time direction.
Its physical meaning is that nothing will escape the spacetime through what is classically 
the singularity. When evaluating the time-symmetric path integral, it amounts to selecting 
$\rho_f$ to be a projection into a pure state $\vert N \rangle$.
Thus, $\rho_f=\vert N\rangle\langle N \vert $. The state $\vert N\rangle$ is that which essentially says 
there is no spacetime beyond what classically would be the singularity and reflects everything.
Of course, this boundary condition only applies to the future inside the black hole. Exterior 
to the black hole, we take the path integral to be controlled by the principal of indifference. 
Alternatively, one can regard the selection of the future final state $\vert N \rangle$ to be a 
boundary condition on solutions of the Wheeler-DeWitt equation. Of course, it remains to be seen 
if the information paradox can be resolved using the time symmetric version 
of the path integral. Were this boundary condition to result in 
some strange behaviour exterior to the black hole, it might either prove to be something
that is detectable leading to a test of these ideas or maybe sufficiently catastrophic to invalidate
the programme. Further investigation of the details of this scenario need to be explored.

One might ask what happens if one insists on having spherically symmetric gravitational 
collapse as opposed to a collapse into a BKL regime. 
The fact that superspace is stratified \cite{Fischer:1970pcv,DeWitt:1969uf} as a result of 
making such symmetry assumptions means that looking here is not going to be relevant to generic
collapse. However, spherically symmetric collapse has been  investigated by Bouhmadi-Lopez {\it et al} 
\cite{Bouhmadi-Lopez:2019kkt}. Their results are not inconsistent with those found here.
 
\section{Unitarity, Causality and the Information Paradox.} 
The quantum picture of a black hole involves the vanishing of the 
gravitational wavefunctional at what would classically be the singularity. If, in addition, one
imposes the condition that the wavefunctional is real, we find
an affirmation of the ideas of Horowitz and Maldacena \cite{Horowitz:2003he}.
The picture we now have has been nicely summarised by Gottesman and Preskill, 
\cite{Gottesman:2003up}.  Particles fall into the black hole as it forms, and then are reflected close
to the singularity by a process that is the time reverse of the Hawking particle-antiparticle 
creation process. Outside the horizon, the usual Hawking 
pair production process can be interpreted as the reflection of antiparticles coming backwards in time 
out of the black hole and scattering into the outgoing Hawking radiation. Quantum information is 
expected to 
follow this flow of particles and is illustrated in Figure $3$. The setting of a 
final boundary condition destroys the usual notion
of causality and unitarity 
and it is this that allows violations of the entropy triangle inequalities and strong
subadditivity \cite{Lloyd:2011zz,Deutsch:1991nm}. 

Bousso and Stanford \cite{Bousso:2013uka}
noted that inside the black hole, it seems that the decoherence functional did not decohere,
leaving the probability interpretation somewhat ambiguous. However we can see this as a failure
of the coarse-graining used.
Whilst such behaviour can be tolerated inside 
the black hole where no observer
is going to make observations that we can perceive, it should not happen outside the black hole.

However, Lloyd \cite{Lloyd:2004wn} observed that as long as normal causality and unitarity 
hold right up to the final surface, classical information is preserved and quantum information escapes
with fidelity $(8/{3\pi})^2$. The classical information is comprised of charges that are embedded
in the classical geometry: the momentum, angular momentum, mass and electric charge of the black hole
together with a collection of soft charges \footnote{Gauge charges other than just electromagnetism 
should presumably be included here too, even though traditionally they 
are not mentioned.}. The practical outcome of Lloyd's result on 
quantum information is that, on average, 
only half a bit of 
quantum information will be lost, independently of the number of bits that escape from the black hole.
Lloyd and Preskill  \cite{Lloyd:2013bza} also showed that this kind of final state model is able to 
avoid the difficulties
presented by firewalls. 

Penrose \cite{Penrose:1969pc} proposed the cosmic censorship hypothesis in order to 
prevent classical singularities being seen by asymptotic observers. The physical reason for
censorship was to prevent singularities influencing physics outside collapsed bodies in an
unpredictable fashion. A slightly different way of expressing these ideas is to ask that the 
Cauchy problem be well-defined when gravitational collapse occurs. Accordingly, singularities 
should always be hidden inside the horizon. There is the possibility that setting future boundary
 conditions inside the horizon could cause violation of the \lq\lq central dogma.\rq\rq\  We suggest 
a quantum cosmic censorship hypothesis, a conjecture that no such trouble occurs. Alternatively,
if quantum cosmic censorship is violated, a weaker version
might well hold, in which such violations occur, 
but are in practice undetectable.

We looked at minisuperspace for pure gravity. One obvious question is to ask what happens if other fields
are included. Something like BKL behaviour seems inevitable. In four spacetime dimensions, ultralocal
BKL type behaviour is almost inevitable. In almost all cases, BKL behaviour and chaos occurs just as 
it does in pure 
gravity, the only difference is that the pattern of the walls changes,
\cite{Belinsky:aa,Damour:2002et,Kleinschmidt:2009cv}. The only exception appears to be if 
there is a massless scalar that does not couple to other fields. If such a scalar exists, then 
the classical approach to singularity is a little more complicated and requires separate treatment. 
It might also be that quantum gravity requires spacetime dimensions greater than 
four in order to control divergences, as is expected to be the 
case in string/M-theory. In spacetime dimensions $d\le 10$, the approach to singularity in pure gravity 
is chaotic. In $d=11$ supergravity the approach to singularity is also chaotic. 
In all of these cases, the wavefunction at the singularity vanishes 
\cite{Kleinschmidt:2009cv} and these wavefunctions can be made real on the approach to singulairty. 
String/M-theory is a promising avenue leading to quantum gravity, but these theories 
(at least in their present form) are not quantum theories of spacetime. However, they seem
to be consistent with the idea that $A_1^{++}$ that controls the approach to singularity in dimension
four is replaced by $E_{10}$, \cite{Damour:2002fz}. It has been speculated that a fundamental theory
of quantum gravity should be based on  $E_{10}$ \cite{Damour:2002cu,Damour:2002et}. If this were the case,
it would add support for our mechanism for avoiding information loss.

Our proposal outlines a plausible route for the complete resolution of the information paradox. 
Further investigation of these ideas is called for, since it seems that the vanishing of the wavefunction
in the future inside a black hole is unavoidable. Precisely how the information is 
retrieved and the placing of these ideas into the broader picture of both gravitational and quantum
physics is an on-going project.

\acknowledgments


I would like to thank the STFC for financial support under grant ST/L000415/1.

I would like to thank Ratindranath Akhoury, David Berman, Sam Braunstein, Jeremy Butterfield, 
Mihalis Dafermos, Fay Dowker, David Garfinkle, Gary Gibbons, 
Hadi Godazgar, Mahdi Godazgar, David Gross, Stephen Hawking, Hermann Nicolai, Frans Pretorius, 
Kip Thorne, Edward Witten and Anna \.Zytkow for stimulating discussions and advice.
 
\vfill\eject
\usetikzlibrary {decorations.pathmorphing}
\vglue -1cm
\begin{tikzpicture}
\draw (0,0) -- (0,14);
\draw (0,0) -- (12,12);
\draw (12,12) -- (4,20);
\draw (0,10) -- (4,14);
\draw [very thick] [decorate, decoration=snake] (0,14) -- (4,14);
\draw (4,14) -- (4,20);
\draw [red, very thick] (12,12) .. controls (8,10) and (2,1)  .. (0,2);
\draw [red, very thick] (12,12) .. controls (8,14)   .. (4,18);
\draw [red, very thick] (0,13.6) .. controls (3.2,13.6) .. (4,14);
\draw [red] (1.4,2.8) node {$\Sigma_i$};
\draw [red] (8,13.8) node {$\Sigma_f$};
\draw [red] (1.6,13.3) node {$\Sigma_s$};
\draw [blue] (-0.4,0.1) node {$i^-$};
\draw [blue] (3.6,19.9) node {$i^+$};
\draw [blue] (12.4,12) node {$i^0$};
\draw [blue] (7,6) node {${\mathcal I}^-$};
\draw [blue] (8.8,16) node {${\mathcal I}^+$};
\draw [blue] (2.5,12) node {${\mathcal H}$};
\draw (1,14.8) node {Singularity};
\end{tikzpicture}
\vskip 1cm
\noindent Figure $1$: 
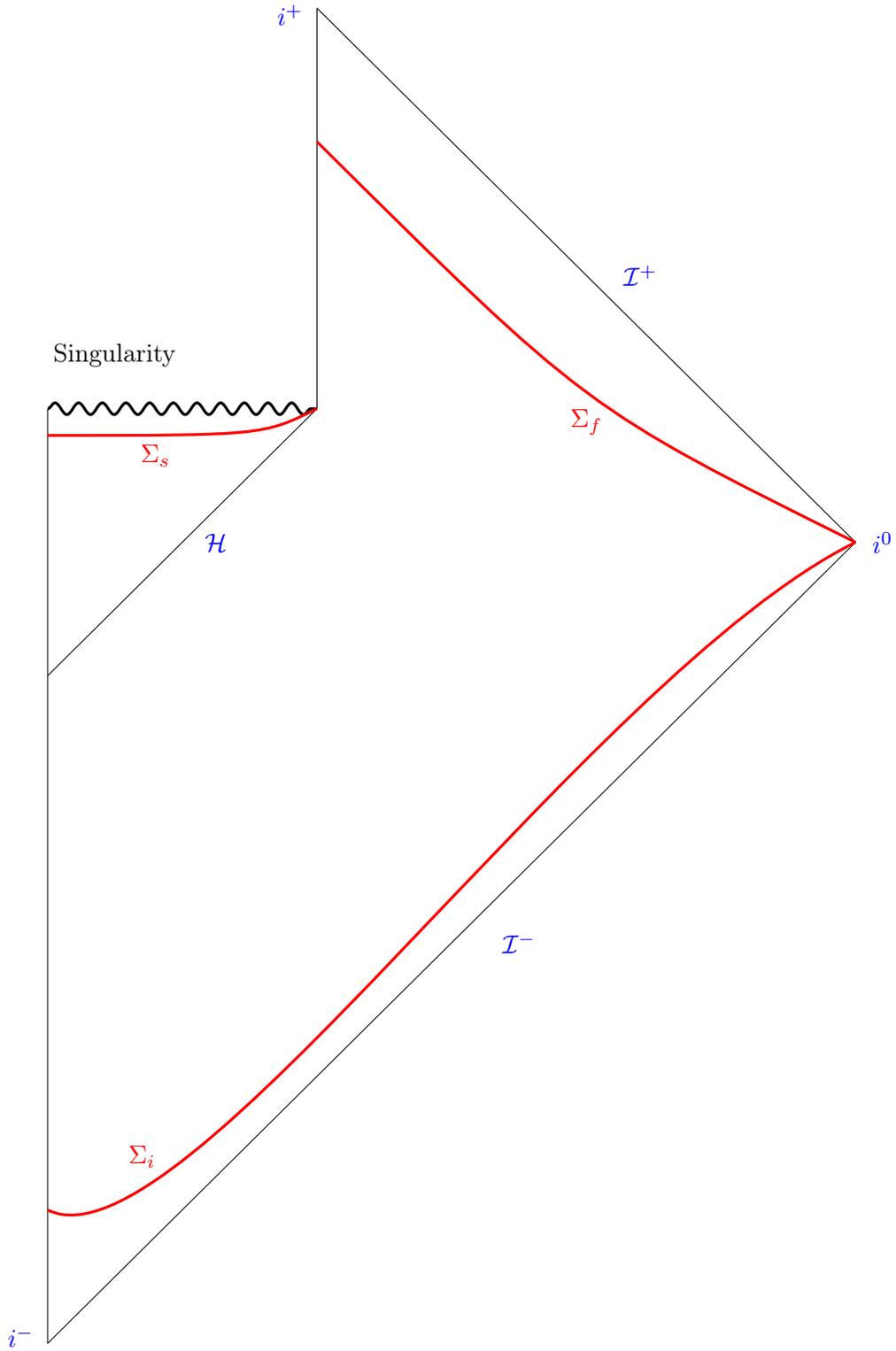
The Penrose diagram for a black hole that evaporates completely. $\Sigma_i$ is an initial Cauchy
surface. $\Sigma_f$ is a similar surface after the black hole has completely evaporated. $\Sigma_s$
is a spacelike surface close to the singularity where future boundary conditions are applied. 
\vfill\eject
\vglue 1cm
\begin{tikzpicture}
\draw (-7.5,0) -- (7.5,0);
\draw (0,-3.0) -- (0,9.0);
\draw (3,0) arc [start angle=0, end angle=180, radius=3];
\draw (1.5,2.58) -- (1.5,9);
\fill[black!20!white] (0,9) -- (1.5,9) -- (1.5,3) -- (0,3) ;
\fill[black!20!white] (1.5,3) -- (1.5,2.58)  
 arc [start angle=60, end angle=90, radius=3] -- (0,3);
\end{tikzpicture}
\vskip 1cm
\noindent Figure $2$: 
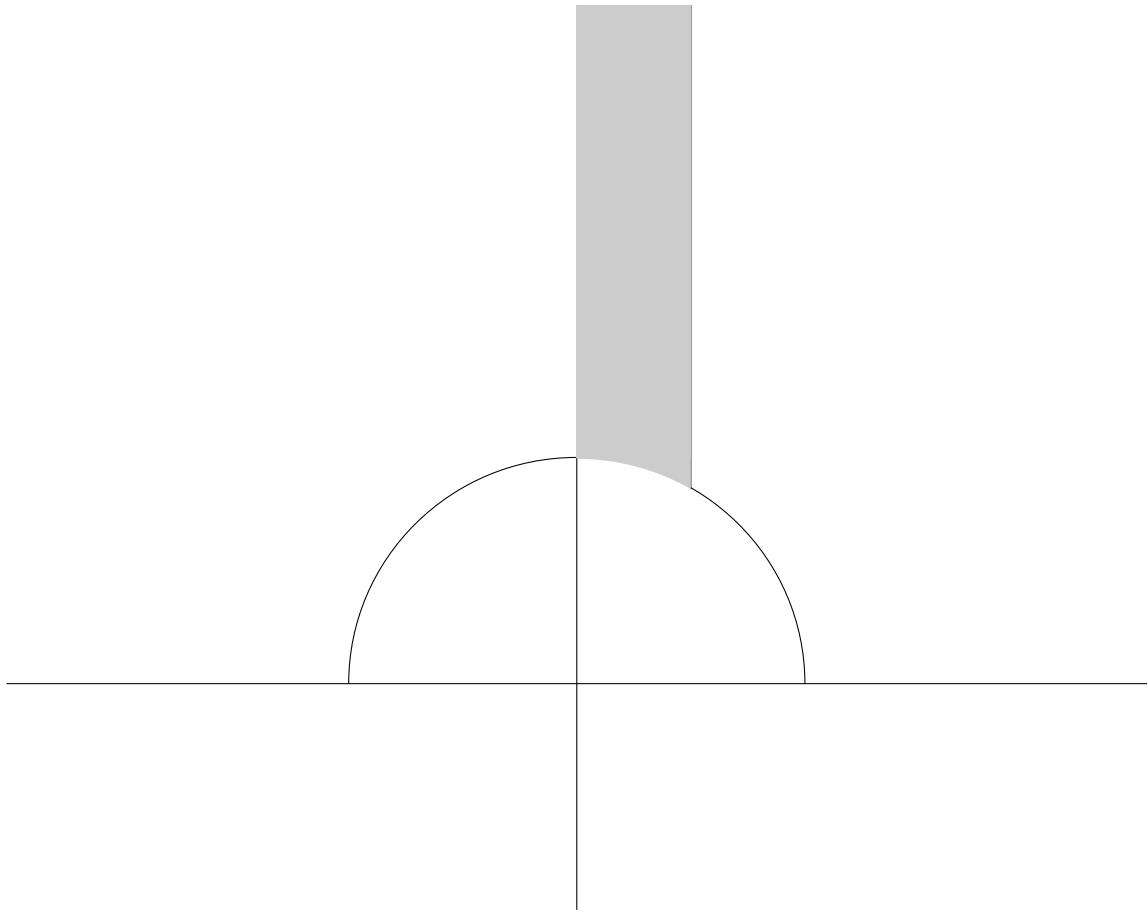
The fundamental region for $PGL(2,\mathbb Z)$ is shaded in grey
and is bounded by the lines 
${\rm Re}\, z=0,\ {\rm Im}\, z\ge 1;\ {\rm Re}\, z=1/2,\ {\rm Im}\, z \ge {\sqrt{3}/}{2}$ and
a segment of the unit circle. 
It is precisely half of the more familiar fundamental  for $PSL(2,\mathbb Z)$.

\vfill\eject

\enlargethispage*{100cm}
\vglue -6cm
\begin{tikzpicture}
\draw (0,0) -- (0,14);
\draw (0,0) -- (12,12);
\draw (12,12) -- (4,20);
\draw (0,10) -- (4,14);
\draw[very thick] [decorate, decoration=snake] (0,14) -- (4,14);
\draw (4,14) -- (4,20);
\draw [blue!20!white, line width=32pt] (1,2) .. controls (1.2,24) and (2.0,5) .. (6,16);
\draw [blue] (-0.4,0.1) node {$i^-$};
\draw [blue] (3.6,19.9) node {$i^+$};
\draw [blue] (12.4,12) node {$i^0$};
\draw [blue] (7,6) node {${\mathcal I}^-$};
\draw [blue] (8.8,16) node {${\mathcal I}^+$};
\draw [blue] (2.5,12) node {${\mathcal H}$};
\end{tikzpicture}
\vskip 1cm
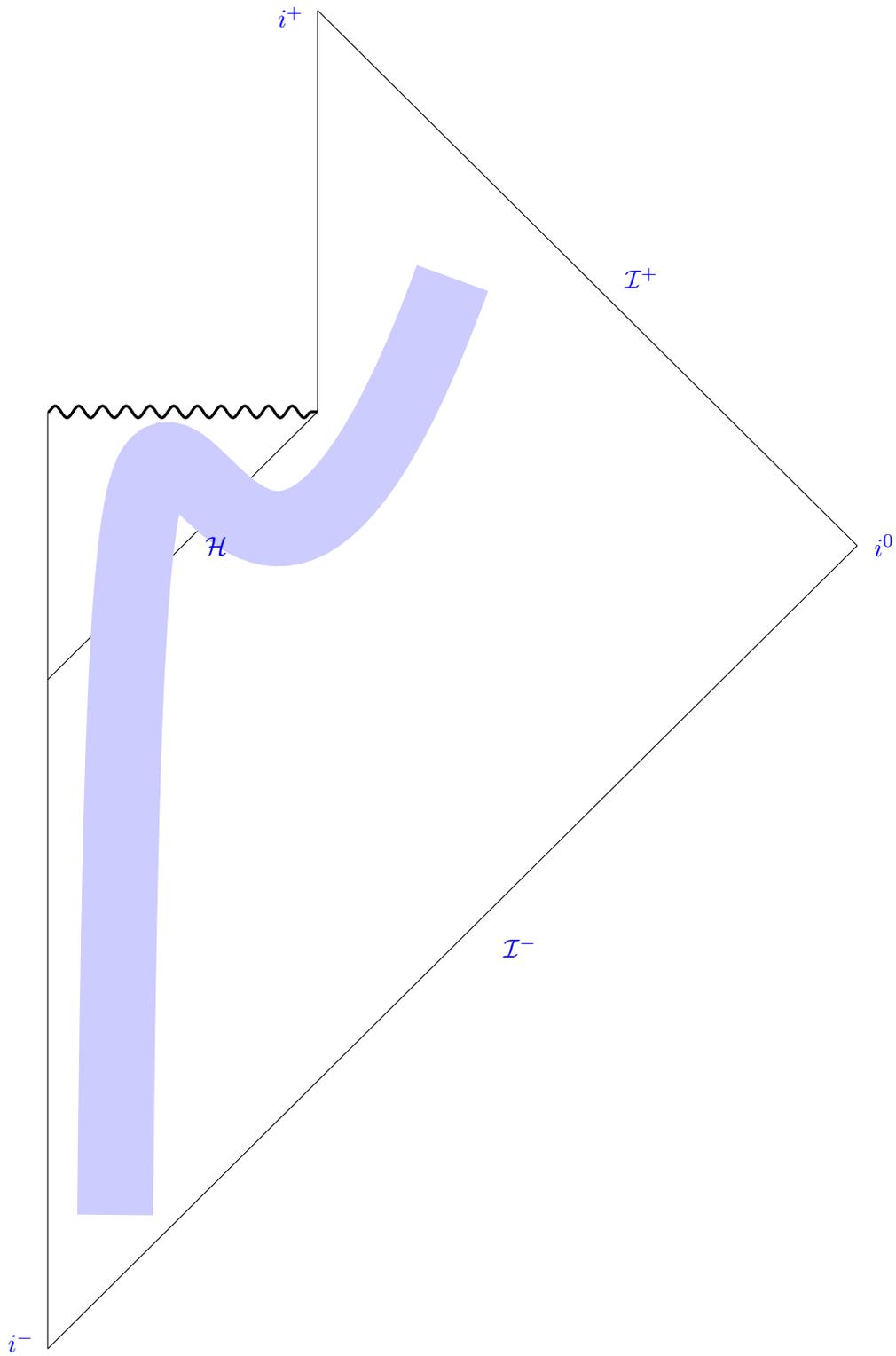
\noindent Figure $3$: The Penrose diagram of a black hole that evaporates completely together
with a river showing the 
expected flow of quantum information.
\vfill\eject

\end{document}